\documentclass[review]{elsarticle}

\usepackage{hyperref}
\usepackage{amssymb}
\usepackage{amsmath}
\usepackage{bm}
\usepackage{multirow}
\usepackage{graphicx}
\usepackage{subfigure}
\usepackage{booktabs}

\journal{Neurocomputing}

\begin{document}

\begin{frontmatter}

\title{RSKNet-MTSP: Effective and Portable Deep Architecture for Speaker Verification}






\author[mymainaddress]{Yanfeng Wu\fnref{firstauthor}}

\author[mysecondaryaddress]{Chenkai Guo\fnref{firstauthor}}
\fntext[firstauthor]{Equal contribution.}
\author[mymainaddress]{Junan Zhao}

\author[mymainaddress]{Xiao Jin}

\author[mymainaddress]{Jing Xu\corref{mycorrespondingauthor}}
\cortext[mycorrespondingauthor]{Corresponding author.}
\ead{xujing@nankai.edu.cn}

\address[mymainaddress]{College of Artificial Intelligence, Nankai University, Tianjin, P.R.China}
\address[mysecondaryaddress]{College of Computer Science, Nankai University, Tianjin, P.R.China}

\begin{abstract}

The convolutional neural network (CNN) based approaches have shown great success for speaker verification (SV) tasks, where modeling long temporal context and reducing information loss of speaker characteristics are two important challenges significantly affecting the verification performance. Previous works have introduced dilated convolution and multi-scale aggregation methods to address above challenges. However, such methods are also hard to make full use of some valuable information, which make it difficult to substantially improve the verification performance. 
To address above issues, we construct a novel CNN-based architecture for SV, called RSKNet-MTSP, where a residual selective kernel block (RSKBlock) and a multiple time-scale statistics pooling (MTSP) module are first proposed. The RSKNet-MTSP can capture both long temporal context and neighbouring information, and gather more speaker-discriminative information from  multi-scale features. In order to design a portable model for real applications with limited resources, we then present a lightweight version of RSKNet-MTSP, namely RSKNet-MTSP-L, which employs a combination technique associating the depthwise separable convolutions with low-rank factorization of weight matrices. Extensive  experiments are conducted on two public SV datasets, VoxCeleb and Speaker in the Wild (SITW). The results demonstrate that 1) RSKNet-MTSP outperforms the state-of-the-art deep embedding architectures by at least 9\%-26\% in all test sets. 2) RSKNet-MTSP-L achieves competitive performance compared with baseline models with 17\%-39\% less network parameters. The ablation experiments further illustrate that our proposed approaches can achieve substantial improvement over prior methods.        

\end{abstract}

\begin{keyword}
speaker verification\sep selective kernel convolution\sep multi-scale aggregation\sep depthwise separable convolution\sep low-rank matrix factorization
\end{keyword}

\end{frontmatter}


\section{Introduction}

\label{sec:introduction}

Speaker verification (SV) aims to verify whether a given utterance belongs to specific speaker according to his/her voice, which is widely used in speech-related fields. To cater to different application scenarios, two typical  categories of SV tasks can be specified, namely  text-dependent SV (TD-SV) and text-independent SV (TI-SV). TD-SV restrains the texts of utterances from being fixed; while TI-SV has no restrictions on the recording utterances. From the implementation algorithms, the SV algorithms can be categorized into stage-wise and end-to-end ones \cite{BAI202165}. The state-wise SV systems consist of a front-end to extract speaker characteristics and a back-end to calculate the similarity score of speaker features. While the end-to-end systems combine the two stages together and calculate the similarity score directly for two input utterances. In this paper, we mainly focus on the state-wise TI-SV systems.

\begin{figure}
    \centering
    \includegraphics[width=\textwidth]{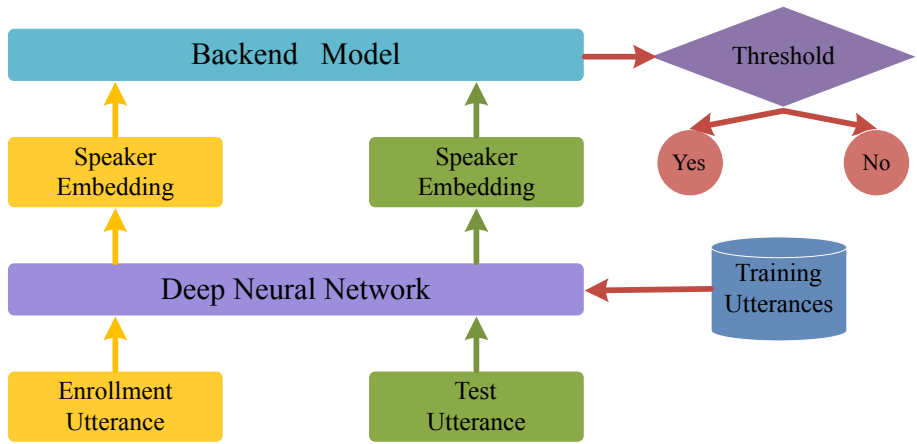}
    \caption{A DNN-based TI-SV system.}
    \label{fig:process}
\end{figure}

In the last few years, the deep learning based approaches have shown great success in SV tasks and achieved significant improvement over the traditional i-vector \cite{5545402} based methods. They adopt deep neural networks (DNNs) to extract speaker features which are also called deep speaker embeddings. A typical DNN-based TI-SV system is shown in Figure \ref{fig:process}, which can be generalized as follows. First, a large amount of utterances are labeled to train a speaker-discriminative DNN architecture. Then, the trained DNN model takes both the enrollment utterance and the test utterance as input, and generates corresponding speaker embeddings respectively.  After that, the two speaker embeddings are fed into a back-end model, such as probabilistic linear discriminant analysis (PLDA) \cite{10.1007/11744085_41} or cosine similarity, to  compute a similarity score. Finally, the verification result can be obtained by comparing the score with a pre-defined threshold.


From the description, the essential part of a DNN-based SV system is to build an effective deep embedding architecture for extracting discriminative features between different speakers.  Currently, the deep embedding architectures can be categorized into two types: frame-level structure and segment-level structure \cite{9203835}. The former one extracts frame-level features for each frame and generate a two-dimensional output with time dimension and channel dimension. Time-delay neural network (TDNN) 
\cite{8461375,8683760,Povey2018,Desplanques2020, WU2020177} and one-dimensional convolutional neural network (CNN) along the time axis, are representative frame-level structures. The segment-level structure \cite{Bhattacharya2017, 8352546, 8683120, BIAN201959, zeinali2019description, XU2020394} consider the input acoustic features as a grayscale image which has three dimensions for time, frequency and channel, respectively, and employ two-dimensional CNN to produce three-dimensional outputs. In segment-level structure, with the downsampling operation, the dimension number of time and frequency dimensions will decrease along with the increase of channel dimensions. After obtaining the outputs, both two types of structures need a pooling layer to aggregate the variable-length features into a fixed-dimensional utterance-level vector. Eventually, the speaker embeddings are extracted from the output of the fully connected (FC) layers.


Due to the limitations of GPU memory and convergence speed, most TI-SV systems use short speech segments of training utterances (a few seconds) to train the DNN while adopt the full-length utterances to extract speaker embeddings, which leads to a mismatch between training and extraction \cite{Garcia-Romero2019}. In addition, since the speech signal is a variable-length sequence in TI-SV, capturing long temporal context with short training segments is a challenging task, which affects the performance of TI-SV systems especially for CNN architectures. Obviously, the CNN units could not interact with the regions outside their receptive fields and are hard to model long context \cite{9053767}. Although some works \cite{Gao2018, MIAO2021201} use dilated convolution to enlarge the receptive fields, they still lose some neighboring information for CNN units, resulting in  performance degradation \cite{8354267}. Another problem in CNN-based models is the information loss of speaker characteristics. As the downsampling operation is applied (e.g. strided convolution), some temporal information is lost due to the reduced time dimensions. To address this problem, multi-scale aggregation methods \cite{Gao2019, Seo2019, Jung2020} are gradually introduced into CNN models. However, these methods usually perform global average pooling (GAP) to aggregate features, leading to information loss of speaker characteristics in both time and frequency dimension.    

On the other hand, with the application for access control in mobile devices \cite{Georges2020}, the design of SV systems tends to be lightweight and efficient. Existing DNN-based SV models, comprising of millions of parameters, require immense computational resources and are hard to achieve the lightweight goal. Thus, it is required for researchers to design more portable models for mobile devices. 



Under above insights, we first propose an effective CNN-based deep embedding architecture called RSKNet-MTSP, which consists of residual selective kernel blocks (RSKBlocks) and a multiple time-scale statistics pooling (MTSP) module. For constructing the RSKBlock, the selective kernel convolution is introduced (SKConv) \cite{Li_2019_CVPR}. Each RSKBlock composes of two SKConv modules and a $1 \times 1$ convolutional layer with residual connections. The SKConv can allow the CNN units to capture both long temporal context and neighbouring information, and provides a soft attention mechanism to adaptively adjust the weight between short and long context. In addition, an MTSP module is designed to further capture the useful speaker information, which extracts multi-scale speaker features in terms of temporal variations over long-term context from multiple layers of the network. 

Based on RSKNet-MTSP, we then propose a lightweight architecture for real applications only with limited resources, called RSKNet-MTSP-L. Compared with RSKNet-MTSP, the RSKNet-MTSP-L adopts two other learning techniques: depthwise seperable convolution and low-rank factorization. 
The depthwise separable convolutions is used to reduce the number of parameters at convolutional layers; while the number of parameters of FC layers is decreased via low-rank matrix factorization of the weight matrices. 

The experiments are conducted on two public TI-SV datasets, VoxCeleb \cite{Nagrani2017,Chung2018} and Speakers in the Wild (SITW) \cite{McLaren+2016}. Experimental results demonstrate that the RSKNet-MTSP achieves superior results than state-of-the-art deep speaker embedding architectures and the RSKNet-MTSP-L also achieves competitive performance compared with baseline models with less network parameters. In addition, the results of ablation experiments further indicate the effectiveness of the components within the constructed architectures. 

The main contributions of this work can be concluded as follows:

\begin{itemize}
    \item [1)] 
    An effective CNN-based deep embedding architecture RSKNet-MTSP is proposed to improve the performance on SV tasks.    
    \item [2)]
    A lightweight architecture RSKNet-MTSP-L is presented  for real applications only with limited resources. 
    \item [3)]
    Extensive experiments conducted on two public SV datasets demonstrate the effectiveness of our proposed architectures. 
\end{itemize}

The remainder of this paper is organized as the follows. Section \ref{sec:relatedwork} describes the related work on DNN-based TI-SV systems. Sections \ref{sec:approach} presents the proposed model. The experimental setup is introduced in Section \ref{sec:experiment}. The results and analysis are shown in Section \ref{sec:result}. Finally, Section \ref{sec:conclusion} concludes the paper.


\section{Related work}

\label{sec:relatedwork}

\subsection{Deep embedding architectures for TI-SV}

In recent years, motivated by the powerful feature extraction capability of DNNs, a lot of DNN-based approaches were proposed and achieved superior performance compared with traditional i-vector \cite{5545402} method. Capturing long temporal context and aggregating variable-length frame-level features serve as two typical challenges in  DNN-based TI-SV systems. 

In order to capture long temporal context, Synder et al. proposed x-vector \cite{Snyder2017,8461375}, exploiting TDNN layers to extract frame-level features. The x-vector has  become the state-of-the-art method for TI-SV because of its excellent performance and light architecture. After that, the TDNN has been widely employed in DNN-based TI-SV architectures (e.g. E-TDNN \cite{8683760}, F-TDNN \cite{Povey2018}, D-TDNN \cite{Yu2020} and ECAPA-TDNN \cite{Desplanques2020}). In addition, lots of works applied well-known CNN structures, e.g., ResNet \cite{8683120, 9054440}, VGGNet \cite{Bhattacharya2017, Yadav2018} and Inception-resnet-v1 \cite{8352546,Li2018} in the extraction of frame-level features. It is still difficult for native CNN to capture long temporal context compared with TDNN. Thus, several studies \cite{Gao2018, MIAO2021201, Jiang2019} introduced dilated convolutions to CNN-based architectures in order to enlarge the receptive fields. Even so, applying too many dilated convolutional layers may lose important neighboring information in deep CNN architecture \cite{8354267}. In this paper, we introduce a selective kernel convolution module \cite{Li_2019_CVPR}, which both takes advantages of both normal and dilated convolution and utilizes a soft attention mechanism to adjust the weights between two kinds of convolutions based on input features. 

Another important factor affecting performance of TI-SV is the pooling layer, which aims to aggregate variable-length frame-level features into a fixed-dimensional vector.  Traditional pooling methods can be generally categorized as temporal pooling and statistics pooling \cite{8461375}. Recently, lots of works introduced attention mechanism into the pooling layer and proposed corresponding attentive pooling methods \cite{Okabe2018, Zhu2018,9053217,Wu2020}. However, previous pooling methods only focus on two axes (i.e., time axis and channel axis) based frame-level features, which is not fit for the outputs of two-dimensional CNN which have three axes (i.e., time axis, frequency axis and channel axis).
On the other hand, prior works presented lots of multi-scale aggregation methods in order to aggregate
speaker information from different time scales in deep CNN architectures \cite{Gao2019, Seo2019, Jung2020}. Nonetheless, these methods usually employ global average pooling, which is hard to make full use of speaker information on time and frequency axis. In this paper, we propose a novel multiple time-scale statistics pooling method, which first converts three-axis features into two-axis features and adopts statistics pooling to capture speaker information in terms of temporal variations over long-term context.



\subsection{Lightweight architectures for TI-SV}

As stated in Section \ref{sec:introduction}, computational limitations restrict the application of DNN-based SV approaches in some advanced scenarios, such as mobile devices. Such challenge motivates researchers to design light and efficient models. Some works introduce model compression methods to address the challenge. Wang et al. \cite{8683443} proposed label-level and embedding-level knowledge distillation (KD) to narrow down the performance gap between large and small CNN-based SV models. Mingote et al. \cite{9053153} explored the KD approach and presented a data augmentation technique to improve robustness of TD-SV systems. Georges et al. \cite{Georges2020} proposed a low-rank
factorized version of the x-vector embedding network \cite{8461375}. To design the lightweight models, Nunes et al. \cite{9207519} proposed a portable model called additive margin MobileNet1D (AM-MobileNet1D) for speaker identification on mobile devices, which uses raw waveform of speeches as input. Safari et al. \cite{Safari2020} presented a deep speaker embedding architecture based on a self-attention encoding and pooling (SAEP) mechanism, which outperforms x-vector \cite{8461375} with less parameters. In this paper, we construct the SV model via two specific lightweight techniques: depthwise separable convolution for reducing the parameters of convolutional layers and low-rank matrix factorization to decreasing the parameters of fully connected layers.   
\section{Architectures}

\label{sec:approach}

\begin{figure}[htb]
    \centering
    \includegraphics[width=\textwidth]{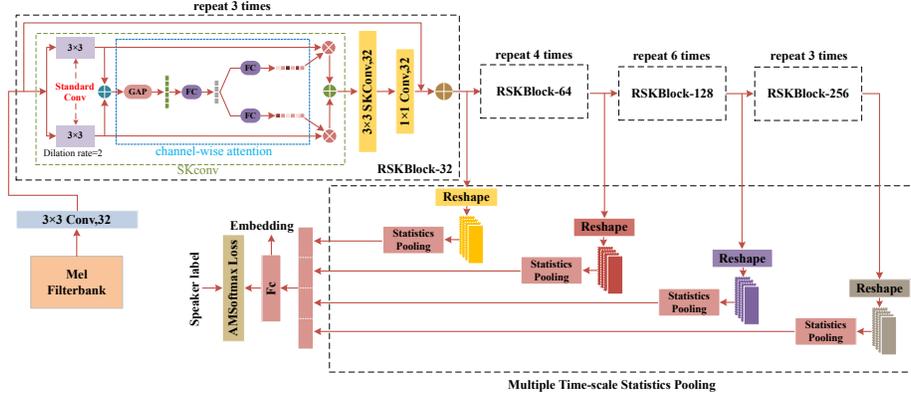}
    \caption{The architecture of RSKNet-MTSP.} 
    \label{fig:rsknetmtsp}
\end{figure}

In this section, we present the two proposed architectures, RSKNet-MTSP and RSKNet-MTSP-L, as shown in Figure \ref{fig:rsknetmtsp} and Figure \ref{fig:rsknetmtspl}, respectively. The two proposed architectures have the same backbone. 

\subsection{RSKNet-MTSP}


The RSKNet-MTSP architecture consists of a $3 \times 3$ convolutional layer, 16 residual selective kernel blocks (RSKBlocks), a multiple time-scale statistics pooling (MTSP) module and an FC layer. The speaker embeddings are extracted from the output of the FC layer, which contains 256-dimensional vectors. In order to obtain more discriminative embeddings, we adopt additive margin softmax (AM-Softmax) loss in the training.

\subsubsection{Residual selective kernel blocks}

Dilated convolution is a kind of special convolution where a filter is employed over an area larger than its length via skipping input values with a certain step. Compared with normal convolution, dilated convolution can obtain a larger receptive field size, which is beneficial to model long-term context for variable-length utterances. However, it is difficult to perform dilated convolutions in deep CNN network. First, performing lots of dilated convolutional layers will result in the ``gridding'' problem and will lose some important neighboring information in higher layers of deep CNN architectures \cite{8354267}. Moreover, for the large number of convolutional layers, it is also challenging to determine which layer(s) to apply the dilated convolutions. 

\begin{figure}
    \centering
    \includegraphics[width=\textwidth]{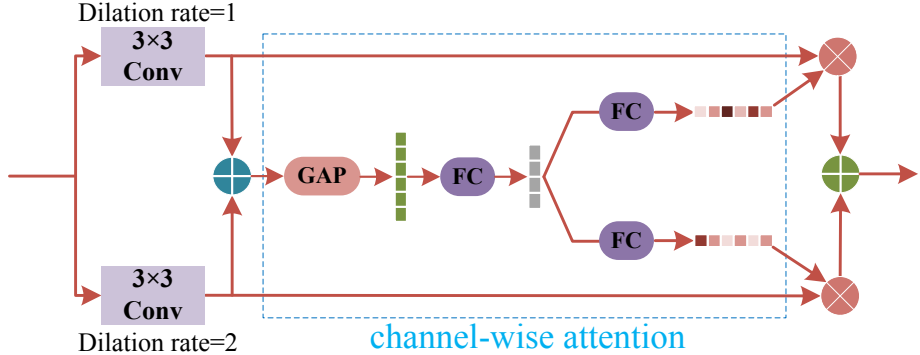}

    \caption{Selective kernel convolution module.}
    \label{fig:skconv}
\end{figure}

To address above challenges, we introduce the selective kernel convolution (SKConv) module  \cite{Li_2019_CVPR} into the deep speaker embedding architecture. The main idea is that standard convolutional layers and dilated convolutional layers are executed on two parallel paths to capture various-length context; they are then fused by a self-attention module to capture different information between long and short context. After that, the SKConv can allow the network to adaptively capture short-term or long-term temporal context based on variable-length input utterances. Such process is shown as illustrated in Figure \ref{fig:skconv}. 

Suppose the input of the SKConv module is $\bm{X} \in \mathbb{R}^{T^{\prime} \times F^{\prime} \times C^{\prime}} $, where $T^{\prime}$,$F^{\prime}$,$C^{\prime}$ are the dimension along the axes of time, frequency and channel, respectively. SKConv performs a $3\times3$ convolution $\widehat{F}$ and a $3\times3$ dilated convolution $\widetilde{F}$ respectively: 
\begin{equation}
\label{nconv}
    \delta(\bm{B}(\widehat{F})): \bm{X} \rightarrow \bm{\widehat{U}} \in \mathbb{R}^{T \times F \times C},
\end{equation}
\begin{equation}
\label{dconv}
    \delta(\bm{B}(\widetilde{F})): \bm{X} \rightarrow \bm{\widetilde{U}} \in \mathbb{R}^{T\times F \times C}.
\end{equation}
In Equation \ref{nconv} and \ref{dconv}, $\bm{B}$ and $\delta$ denote batch normalization (BN) and ReLU function, respectively. 
Then an element-wise summation is applied to fuse features from two branches: 
\begin{equation}
\label{getu}
   \bm{U} = \bm{\widehat{U}} \oplus \bm{\widetilde{U}},
\end{equation}
where $\oplus$ denotes the element-wise summation; $ \bm{U}$ is the global representation obtained from two paths. 

Then a soft attention mechanism across the channel axis is employed upon the global feature $\bm{U}$, and a global average pooling (GAP) operation is performed to generate channel-wise vector $\bm{s} \in \mathbb{R}^{C}$: 
\begin{equation}
    s_c = \frac{1}{T \times F}\sum^{T}_{t=1}\sum^{F}_{f=1} \bm{U}_c(t,f), 
\end{equation}
where $c$ is the index of the channel axis. Further, a hidden layer is applied to generate a compact feature $\bm{z}$ with the reduction of dimension:
\begin{equation}
    \bm{z} = \delta(\bm{B}(\bm{W} \bm{s})),
\end{equation}
where $\bm{W} \in \mathbb{R}^{ g \times C}$ is the weight matrix of the hidden layer, and $g$ is defined as follows:
\begin{equation}
    g = \max(\frac{C}{r}, L),
\end{equation} 
where $r$ is the reduction ratio and $L$ is set to prevent $g$ being too small. In practical implementation, $r$ is set to 16 and $L$ is set to 32.

Finally, an FC layer followed by a softmax function is executed to obtain the channel-wise attentive vector for features from each path respectively: 
\begin{equation}
    a_c = \frac{e^{\mathbf{A}_{c}\mathbf{z}}}{e^{\mathbf{A}_{c}\mathbf{z}}+e^{\mathbf{B}_{c}\mathbf{z}}},b_c = \frac{e^{\mathbf{B}_{c}\mathbf{z}}}{e^{\mathbf{A}_{c}\mathbf{z}}+e^{\mathbf{B}_{c}\mathbf{z}}},
\end{equation}
where $\mathbf{A}, \mathbf{B} \in \mathbb{R}^{C \times g}$ denote the weight matrices of the FC layers for $\widehat{U}$ and $\widetilde{U}$, respectively. $\mathbf{A}_{c}, \mathbf{B}_{c} \in \mathbb{R}^{1 \times g} $ is the $c$-th row of $\mathbf{A}$ and $\mathbf{B}$, respectively. The softmax operator ensures that the sum of all elements on each channel is equal to 1.

Finally, the output feature $\bm{V}$ of SKConv module is the weighted element-wise summation of features on two paths across the channel axis: 
\begin{equation}
    \bm{V}_c = \widehat{a}_c \cdot \bm{\widehat{U}}_c \oplus \widetilde{b}_c \cdot \bm{\widetilde{U}}_c,
\end{equation}
where $\bm{V} = [\bm{V}_1,\bm{V}_2,...,\bm{V}_C] $ and $\bm{V}_c \in \mathbb{R}^{T \times F}$. 

In this paper, we propose a residual SKConv Block (RSKBlock), which performs SKConv with residual connections \cite{He_2016_CVPR}. The residual block includes two kind of blocks, the building block and the bottleneck block. The bottleneck block expands the channel dimension of output features, and thus increases the model complexity. Hence, the RSKBlock is designed as the building block. Since the output of SKConv module has been processed by the non-linear ReLU function, it can not be used for residual connections. We add a $1 \times 1$ convolutional layer after SKConv modules to solve this problem. As shown in Figure \ref{fig:rsknetmtsp}, the input of RSKBlocks are fed into two SKConv modules, a $1 \times 1$ convolutional layer, a batch normalization layer, a residual connection and a ReLU function. Inspired by the ResNet structure, we construct the RSKNet architecture which contains four RSK modules by 3, 4, 6 and 3 RSKBlocks, respectively. 

\subsubsection{Multiple time-scale statistics pooling}

Since the length of the input feature of the network is variable on time axis, most deep speaker embedding models employ a pooling layer to map the variable-length frame-level features to a utterance-level fixed-dimensional embedding. Deep CNN networks usually contains several downsampling layers which reduce the spatial size of feature maps and enlarge the number of channels. These downsampling operations can be implemented by max pooling layers or convolutional layers with a stride greater than 1. Thus, deep CNN networks can produce features of different time-frequency scales and resolutions. The features from higher layers have less time frames but each frame has more speaker discriminative information. In DNN-based TI-SV, the pooling layer captures speaker characteristics from all frames and aggregates them to generate a embedding vector. The aggregated multi-scale features can gather information from more time frames. In this section, we introduce the proposed multiple time-scale statistics pooling method.  In r-vectors \cite{zeinali2019description}, a statistics pooling method is proposed for CNN-based features and shows superior performance.  Since statistics pooling can capture the speaker’s characteristics in terms of the temporal variations over long-term context, we present its multi-scale version to collect various temporal and frequency information from multiple time-scale features, as illustrated in Figure \ref{fig:rsknetmtsp}. 

We denote the output of each residual module as $\bm{X}_i \in \mathbb{R}^{T_i \times F_i \times C_i} $ where  $i=1,2,3,4$ and $T_i, F_i, C_i$ denote the dimension of time, frequency and channel axes, respectively. We first converts the three-axis features into two-axis features:
\begin{equation}
    \bm{X}_i \in \mathbb{R}^{T_i \times F_i \times C_i} \rightarrow \bm{X}_{i}^{\prime} \in \mathbb{R}^{T_i \times N_i}, \forall i\in \{1,2,3,4\}, 
\end{equation}
where $N_i = F_i \cdot C_i$. Then, a statistics pooling operation along time axis is applied to compute a mean feature vector $\bm{\mu}_i \in \mathbb{R}^{N_i} $  and a standard deviation feature vector $\bm{\sigma}_i \in \mathbb{R}^{N_i} $ for each output:
\begin{equation}
    \mu_{i}(n) = \frac{1}{T} \sum^{T}_{t=1} \bm{X}_{i}^{\prime}(t,n), \forall i\in \{1,2,3,4\},
\end{equation}
\begin{equation}
    \sigma_{i}(n) = \sqrt{\frac{1}{T}\sum^{T}_{t=1}\bm{X}_{i}^{\prime}(t,n)^2-\mu_{i}(n)^2}, \forall i\in \{1,2,3,4\}, 
\end{equation}
where $t, n$ represent the time index and the channel index of $\bm{X}_{i}^{\prime}$, respectively. Finally, we concatenate the feature vectors obtained from all residual modules to acquire the output of the statistics pooling layer:
\begin{equation}
    \bm{e} = [\bm{\mu}_{1},\bm{\sigma}_{1},...,\bm{\mu}_{i},\bm{\sigma}_{i}], \forall i\in \{1,2,3,4\}.
\end{equation}  

\subsubsection{Additive margin softmax loss}

Traditional cross-entropy based softmax loss is widely used in classification tasks but do not perform well on producing discriminative features between different speakers. Additive margin softmax (AM-Softmax) loss is first proposed for face verification \cite{8331118} , and then extended to a more discriminative model for SV \cite{Liu2019}. AM-Softmax loss is calculated as follows:
\begin{equation}
\label{eq:amloss}
L=-\frac{1}{N}\sum^{N}_{i=1} \log \frac{e^{s(\cos{\theta_{y_i}}-m)}}
{e^{s(\cos{\theta_{y_i}}-m)}+\sum^{C}_{j=1,j\neq y_i}e^{s\cos{\theta_{j}}}}.
\end{equation}
In Equation \ref{eq:amloss}, $N$ is the size of a batch; $C$ is the number of speakers in the training set; $y_i$ is the ground truth label of the $i$-th sample. The scaling factor $s$ and the margin $m$ are two hyperparameters for AM-Softmax loss. In this paper, we use the AM-Softmax loss to train the proposed models for better efficiency of TI-SV tasks. 

\subsection{RSKNet-MTSP-L}

\begin{figure}[htb]
    \centering
    \includegraphics[width=\textwidth]{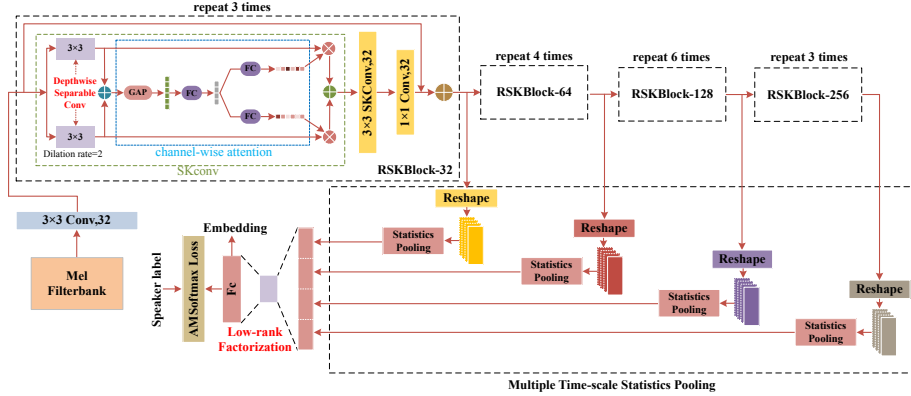}
    \caption{The architecture of RSKNet-MTSP-L.} 
    \label{fig:rsknetmtspl}
\end{figure}

Although the RSKNet-MTSP can achieve better TI-SV performance than several baseline models, which can be verified in Section \ref{sec:result}, it has more network parameters. For instance, the number of parameters of RSKNet-MTSP is 2.3 times that of the ResNet34-SP model. In order to design a portable model for real applications with limited resources, we build a lightweight architecture, RSKNet-MTSP-L, based on RSKNet-MTSP, as shown in Figure \ref{fig:rsknetmtspl}. Since most parameters of the network are from convolutional layers and FC layers, compared with RSKNet-MTSP, RSKNet-MTSP-L employs depthwise separable convolution instead of standard convolution and performs low-rank factorization on the weight matrix of the FC layer to reduce model size.

\subsubsection{Depthwise separable convolution}

\begin{figure}
    \centering
    \includegraphics[width=\textwidth]{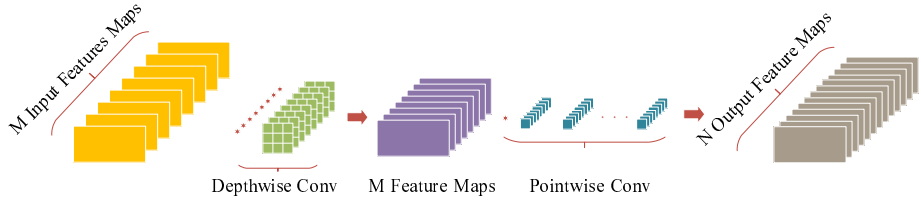}
    \caption{Depthwise separable convolution.}
    \label{fig:dwconv}
\end{figure}

The depthwise separable convolution is inspired by group convolution and inception modules, and is widely employed in the networks for image classification, such as MobileNet \cite{howard2017mobilenets,Sandler_2018_CVPR} and Xception \cite{Chollet_2017_CVPR}, which is typically composed of two parts: a depthwise convolution and a pointwise convolution, as illustrated in Figure \ref{fig:dwconv}. The depthwise convolution is a spatial convolution with a single filter applied independently on each channel of the input feature. Thus, the number of output channels is equal to that of input channels. Next, a pointwise convolution with $1 \times 1$ kernel projects the channels computed by the depthwise convolution onto the new channels. Pointwise convolution changes the number of channels but does not change the spatial size of the feature. 

The parameters of the depthwise separable convolution are significantly less than those of the standard convolution. Suppose that the spatial size of the convolutional kernel is $j \times k$, and the channel number of input and output is $i$ and $o$, respectively. The parameters of standard convolution is $j\cdot k \cdot i \cdot o$, and the parameters of depthwise separable convolution is $j \cdot k \cdot i+i \cdot o$. The ratio of the parameter of two convolutions is denoted as:
\begin{equation}
    \frac{Param_{DSC}}{Param_{SC}} = \frac{j \cdot k \cdot i+i \cdot o}{j \cdot k \cdot i \cdot o} = \frac{1}{o}+\frac{1}{j \cdot k}.
\end{equation}

In original SKConv modules \cite{Li_2019_CVPR}, group convolution is applied in convolutional layers and shows competitive performance. Thus, in RSKNet-MTSP-L, we adpot depthwise separable convolution in convolutional layers of all RSKBlocks. 


\subsubsection{Low-rank factorization}

\label{sec:lowrank}

Although the MTSP module gathers the speaker information from different time scales, it generates a high-dimensional feature vector $\bm{e}$. For example, MTSP produces a output vector which has 10240 dimensions in our implementation, leading to a large weight matrix of the following FC layer. In order to address the problem, we use low-rank matrix factorization to reduce the parameters of the FC layer, as shown in Figure \ref{fig:rsknetmtspl}.

Suppose that the output of the FC layer is $\bm{y} \in \mathbb{R}^{n}$, and the input $\bm{e}$ has $m$ dimensions. The FC layer can be defined as follows: 
\begin{equation}
    \bm{y} = \bm{W_{1}} \bm{e} + \bm{b},
\end{equation}
where $\bm{W_{1}} \in \mathbb{R}^{n \times m}, \bm{b} \in \mathbb{R}^{n} $ denotes the weight matrix and bias of the FC layer, respectively; the number of weights is $n \cdot m$. With the low-rank matrix factorization, the matrix described above is replaced by two matrices $\bm{W}_{2} \in \mathbb{R}^{n \times p} $ and $\bm{W}_{3} \in \mathbb{R}^{p \times m} $, where $p$ is the low-rank constant and $1<p<min(m,n)$. The output of the FC layer can be calculated as follows:
\begin{equation}
    \bm{y} = \bm{W}_{2} \bm{W}_{3} \bm{e} + \bm{b}.
\end{equation}
After the low-rank factorization, the number of weights $n \cdot p + p \cdot m$ decreases significantly only if $p$ is set properly.  In the experiment, we will further evaluate verification performance and model parameters under the different value of $p$ to explore the suitable $p$ value.

\section{Experimental setup}

\label{sec:experiment}

\subsection{Dataset}

The experiments are conducted on two well-known public speaker recognition datasets: VoxCeleb \cite{Nagrani2017,Chung2018} and Speaker in the Wild (SITW) \cite{McLaren+2016}. VoxCeleb is a large speaker recognition dataset gathered from open-source media, which contains over a million utterances, 7,000 speakers and 2000 hours. For VoxCeleb dataset, the average duration of utterances is 8 seconds and most utterances are short utterances with the duration less than 10 seconds. VoxCeleb includes two sub datasets, VoxCeleb1 and VoxCeleb2. The SITW dataset contains recordings of 299 public celebrities from open-source media. The length of speech segments in SITW ranges from 6 seconds to 180 seconds and most segments are long utterances. Hence, the two datasets can be used to evaluate the performance of our proposed architectures on utterances with various lengths. 

\begin{table}[htb]
    \centering
    \caption{Datasets used in the experiments. Note that the speaker number is not listed here since partial speaker models in SITW belong to same speakers. }
    \begin{tabular}{ccccc}
    \toprule
    Split & Dataset & Speakers & Utterances & Trials \\
    \midrule
    Training & VoxCeleb2-Dev & 5,994 & 1,092,009 & - \\
    \hline
    \multirow{5}{*}{Test} & VoxCeleb1-O & 40 & 4,706 & 37,611 \\
    & VoxCeleb1-E & 1,251 & 142,540 & 579,818 \\
    & VoxCeleb1-H & 1,190 & 135,415 & 550,894 \\
    & SITW-Dev & - & 823 & 338,226 \\
    & SITW-Eval & - & 1,202 & 72,1788 \\
    \bottomrule
    \end{tabular}
    \label{tab:dataset}
\end{table}

Each of the three datasets, VoxCeleb1, VoxCeleb2 and SITW, are divided into a development part and a test (evaluation) part. We choose the development part of VoxCeleb2 (VoxCeleb2-Dev) as the training set, which contains 1,092,009 utterances and 5,994 speakers. The rest of the datasets as treated as test sets, including three parts: whole VoxCeleb1 dataset, SITW development set and SITW evaluation set. 
The VoxCeleb1 has three different SV tasks: original task (VoxCeleb1-O), extended task (VoxCeleb1-E) and hard task (VoxCeleb1-H), which are systematically evaluated in our experiments. Both SITW development set (SITW-Dev) and SITW evaluation (SITW-Eval) set have four different SV tasks: core-core, core-multi, assist-core, assist-multi. We apply the core-core task for evaluation, in which both enrollment speech segments and test speech segments are derived from only a single speaker. The dataset details are listed in Table \ref{tab:dataset}.

\subsection{Baseline models}


\begin{table}[htb]
    \centering
    \caption{Different structures of ResNet34. $r$ denotes dilation rate.}
    \resizebox{\textwidth}{!}{
    \begin{tabular}{ccccc}
    \toprule
    Layer name & ResNet34 & DResNet34-1 & DResNet34-2 \\
    \midrule
    Conv2D  & $3 \times 3, 32$ & $3 \times 3, 32$ & $3 \times 3, 32$ \\
    ResBlocks-1 & $\left[\begin{array}{l}{3 \times 3,32} \\ {3 \times 3,32}\end{array}\right] \times 3$ & $\left[\begin{array}{l}{3 \times 3,32,r=2} \\ {3 \times 3,32,r=2}\end{array}\right] \times 3$ & $\left[\begin{array}{l}{3 \times 3,32} \\ {3 \times 3,32}\end{array}\right] \times 3$  \\
    ResBlocks-2 & $\left[\begin{array}{l}{3 \times 3,64} \\ {3 \times 3,64}\end{array}\right] \times 4$ & $\left[\begin{array}{l}{3 \times 3,64,r=2} \\ {3 \times 3,64,r=2}\end{array}\right] \times 4$ & $\left[\begin{array}{l}{3 \times 3,64} \\ {3 \times 3,64}\end{array}\right] \times 4$  \\
    ResBlocks-3 & $\left[\begin{array}{l}{3 \times 3,128} \\ {3 \times 3,128}\end{array}\right] \times 6$ & $\left[\begin{array}{l}{3 \times 3,128,r=2} \\ {3 \times 3,128,r=2}\end{array}\right] \times 6$ & $\left[\begin{array}{l}{3 \times 3,128,r=2} \\ {3 \times 3,128,r=2}\end{array}\right] \times 6$ \\
    ResBlocks-4 & $\left[\begin{array}{l}{3 \times 3,256} \\ {3 \times 3,256}\end{array}\right] \times 3$ & $\left[\begin{array}{l}{3 \times 3,256,r=2} \\ {3 \times 3,256,r=2}\end{array}\right] \times 3$ & $\left[\begin{array}{l}{3 \times 3,256,r=4} \\ {3 \times 3,256,r=4}\end{array}\right] \times 3$  \\
    \bottomrule
    \end{tabular}
    }
    \label{tab:resnet}
\end{table}

In order to evaluate the performance of the proposed models, we have implemented four state-of-the-art deep speaker embedding architectures: TDNN \cite{8461375}, E-TDNN \cite{8683760}, ECAPA-TDNN \cite{Desplanques2020} and ResNet34-SP \cite{zeinali2019description}. TDNN, E-TDNN and ECAPA-TDNN are frame-level structures and ResNet34-SP belongs to segment-level structure. 

TDNN, also called x-vector, is a TDNN-based deep embedding baseline architecture. The frame-level structure contains five TDNN layers followed by ReLU and Batch Normalization (BN). Then, it employs a statistics pooling (SP) layer to aggregate the frame-level outputs into a fixed-dimensional vector. The outputs of SP layer are fed into two FC layers for extracting speaker embeddings. In our implementation, the speaker embedding is extracted from the output of first FC layer. The detailed network of x-vector can be found in \cite{8461375}. 

E-TDNN can be treated as an improved version of x-vector. Compared to x-vector, E-TDNN utilizes a wider temporal context in the TDNN layers, and adds dense layers between the TDNN layers. Generally, E-TDNN have more parameters as well as better performance than x-vector. The network configuration of E-TDNN can be found in \cite{8683760}. 

ECAPA-TDNN is proposed to improve the performance of TDNN-based TI-SV systems, with multiple advanced techniques such as one-dimensional Res2Net modules, residual connections, Squeeze-and-Excitation blocks, and channel-dependent statistics pooling. The channel dimension in the convolutional frame layers is set to 512 and other settings is as same as \cite{Desplanques2020}. 

ResNet34-SP is a very deep two-dimensional CNN  combined with both ResNet34 topology and the SP layer. In ResNet34-SP, the three-dimensional features are reshaped into two-dimensional features with time and channel axes before fed into the SP layer. Such architecture achieves the best performance on TI-SV tasks in VoxCeleb Speaker Recognition Challenge (VoxSRC) 2019. The structure of ResNet34 is shown in Table \ref{tab:resnet}.

\subsection{Implementation details} 

We employ the Kaldi toolkit \cite{Povey_ASRU2011} to handle the data processing step, which does not contain any augment handling for the training data. The input features of the TDNN-based models are 80-dimensional Mel frequency cepstral coefficients (MFCCs), while the input features of the CNN-based models are 40-dimensional Mel filterbanks (FBank) features for CNN-based models. Both kinds of acoustic features have a frame length of 25ms and a frame shift of 10ms. The energy-based voice activity detection (VAD) and cepstral mean normalization with a 3-second window are applied to the acoustic features. In the training process, the utterances are cut into 2-second segments (corresponding to 200 frames), while in the test, the entire utterances are fed into the network to extract corresponding speaker embeddings. 

We adopt an existing Pytorch toolkit \cite{9414676} to conduct the network training and feature extraction. All networks are trained with the batch size of 128 and optimized by stochastic gradient descent with the momentum of 0.9. We train all models with AM-Softmax loss for fair comparison, and $m$ and $s$ are set to 0.2 and 30, respectively. The learning rate is initialized as 0.01 and then is continuously divided by 10 once the validation loss fails to decrease for a while, till it reaches 1e-6. 

We follow the same backend as the implementation of baseline systems. The similarity scores are computed with PLDA for TDNN-based models, and with cosine distance for CNN-based models. The extracted embeddings are centered by subtracting the mean of embeddings, which is computed by all training utterances and then length-normalized. For the PLDA backend, we apply a linear discriminant analysis (LDA) to reduce the dimensionality of embeddings in TDNN and E-TDNN from 512 to 160. 


\subsection{Evaluation metric}

In our experiments, the performance of all TI-SV systems are evaluated in terms of equal error rate (EER) and the minimum of normalized detection cost function (MinDCF) \cite{Sadjadi2017}. The MinDCF is calculated as the a weighted sum of false-reject (FR) and false-accept (FA) error probabilities:
\begin{equation}
\label{eq:det} 
\begin{aligned}
    C_{Det} = & C_{FR} \times P_{Target} \times P_{FR|Target} + \\
    & C_{FA} \times (1-P_{Target}) \times P_{FA|NonTarget},
\end{aligned}
\end{equation}
and then normalized:
\begin{equation}
    C_{Norm} = \frac{C_{Det}}{C_{Default}},
\end{equation}
where $C_{Default}$ is defined as follows:
\begin{equation}
    C_{Default} = \min(C_{FR} \times P_{Target},  C_{FA} \times (1-P_{Target})). 
\end{equation}
In Equation \ref{eq:det}, $C_{FR}$ is the cost of missed detection; $C_{FA}$ is the cost of a spurious detection; $P_{Target}$ is a prior probability of the specified target speaker. In this paper, we set $C_{FR}=C_{FA}=1$ and $P_{Target}=0.01$, which is commonly used in relevant experiments \cite{Chung2018, Jung2020}.

\section{Result}

\label{sec:result}

\subsection{Overall performance}

\begin{table}[htb]
    \centering
    \caption{Results for various models in terms of EER and MinDCF on SITW. }
    \begin{tabular}{cccccc}
    \toprule
    \multirow{2}{*}{Model} & \multirow{2}{*}{Params} & \multicolumn{2}{c}{SITW-Dev} & \multicolumn{2}{c}{SITW-Eval}  \\
    \cline{3-6}
    &  & EER & MinDCF & EER & MinDCF \\
    \midrule
    TDNN & 4.6M  & 3.24  & 0.314 & 3.53 & 0.349   \\
    E-TDNN & 6.8M  & 2.77 & 0.292  & 3.25 & 0.319   \\
    ECAPA-TDNN & 6.2M & 2.19 & 0.257 & 2.82 & 0.289   \\
    ResNet34-SP & 6.0M & 2.27 & 0.217 & 2.82  & 0.244  \\
    \textbf{RSKNet-MTSP} & 13.9M & \textbf{1.85} & \textbf{0.183} & \textbf{2.27} & \textbf{0.228}    \\
    \textbf{RSKNet-MTSP-L} & 3.8M & 2.46 & 0.225  & 2.84 & 0.264  \\ 
    \bottomrule
    \end{tabular}
    \label{tab:sitw}
\end{table}

Table \ref{tab:sitw} and Table \ref{tab:voxceleb} list the results in terms of EER and MinDCF on VoxCeleb1 and SITW respectively, containing four baseline models and our proposed two models.  There are three tasks in VoxCeleb1, and two tasks in SITW. From Table \ref{tab:sitw}, it can be observed that ECAPA-TDNN and ResNet34-SP significantly outperform TDNN and E-TDNN. In the two tasks of SITW, RSKNet-MTSP achieves the best performance on both EER and MinDCF. Compared to ResNet34-SP, RSKNet-MTSP improves 18.5\% for SITW-Dev and 19.5\% for SITW-Eval in terms of EER. However, RSKNet-MTSP has more network parameters than the baseline models, costing more GPU memory. To address the resource cost, we propose the lightweight model  RSKNet-MTSP-L, which only contains 3.8M parameters, less than all baseline models. It reduces the network parameters of 17.4\%, 38.7\%, 38.7\% and 36.7\% compared with TDNN, E-TDNN, ECAPA-TDNN and ResNet34-SP. In Table \ref{tab:sitw}, the RSKNet-MTSP-L performs subtle poorly compared to RSKNet-MTSP, indicating that parameter reduction weakens the modeling capabilities. However, it achieves superior performance than TDNN and E-TDNN, and have similar EERs with ECAPA-TDNN and ResNet34-SP on SITW-Eval, demonstrating its effectiveness.   

\begin{table}[htb]
    \centering
    \caption{Results of the models on VoxCeleb.}
    \begin{tabular}{ccccccc}
    \toprule
    \multirow{2}{*}{Model} & \multicolumn{2}{c}{VoxCeleb1-O} & \multicolumn{2}{c}{VoxCeleb1-E} & \multicolumn{2}{c}{VoxCeleb1-H} \\
    \cline{2-7}
    & EER & MinDCF & EER & MinDCF & EER & MinDCF \\
    \midrule
    TDNN & 2.13 & 0.246 & 2.27  & 0.251  & 3.86 & 0.373 \\
    E-TDNN & 2.00 & 0.268 & 2.19 & 0.250 & 3.75 & 0.369 \\
    ECAPA-TDNN & 1.55 & 0.198 & 1.78  & 0.202  & 3.03 &  0.292 \\
    ResNet34-SP & 1.43 & \textbf{0.159} & 1.59  & 0.176  & 2.76 & 0.253 \\
    \textbf{RSKNet-MTSP} & \textbf{1.05} & \textbf{0.159} & \textbf{1.30}  & \textbf{0.152}  & \textbf{2.52} & \textbf{0.237} \\ 
    \textbf{RSKNet-MTSP-L} & 1.68 & 0.173  & 1.71  & 0.180 & 3.00 & 0.272 \\
    \bottomrule
    \end{tabular}
    \label{tab:voxceleb}
\end{table}

The results on VoxCeleb have similar trends for SITW, where RSKNet-MTSP achieves the best EERs and MinDCFs for all three tasks. Specifically, RSKNet-MTSP yields 26.6\%, 18.2\% and 8.7\% improvements in terms of EER compared with ResNet34-SP. RSKNet-MTSP-L generally performs better than ECAPA-TDNN, but worse than ResNet34-SP. However, note that ResNet34-SP is the best model in VoxSRC 2019, which also uses VoxCeleb1 as the test set. RSKNet-MTSP-L is only worse than the best model and achieves superior performance than other baselines models with less network parameters. 

\begin{figure}[htbp]
\centering
\subfigure[SITW-Dev]{
\includegraphics[width=0.45\textwidth]{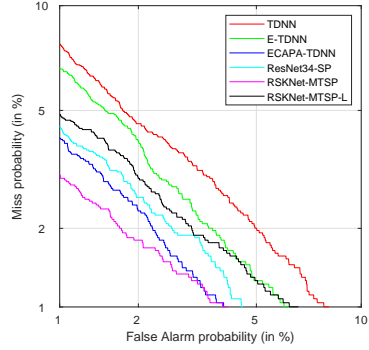}
\label{fig:sitwdev}
}
\quad
\subfigure[SITW-Eval]{
\includegraphics[width=0.45\textwidth]{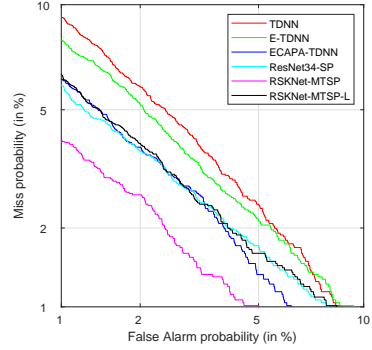}
\label{fig:sitweval}
}
\quad
\subfigure[VoxCeleb1-O]{
\includegraphics[width=0.45\textwidth]{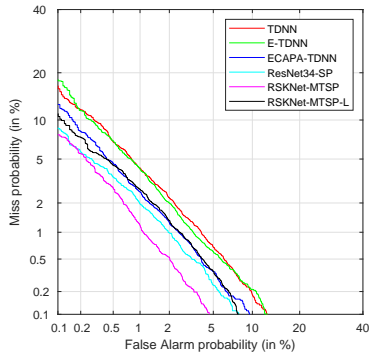}
\label{fig:vox1o}
}
\quad
\subfigure[VoxCeleb1-E]{
\includegraphics[width=0.45\textwidth]{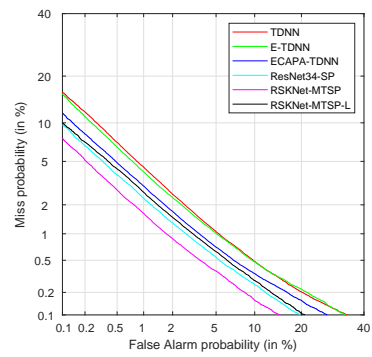}
\label{fig:vox1e}
}
\quad
\subfigure[VoxCeleb1-H]{
\includegraphics[width=0.45\textwidth]{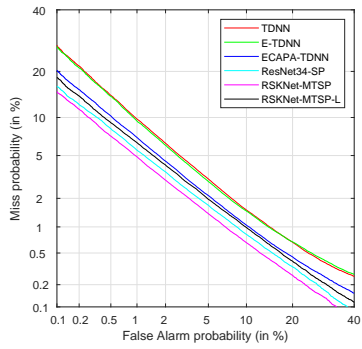}
\label{fig:vox1h}
}
\caption{DET curves for models in different test sets.}
\label{fig:det}
\end{figure}

To further visualize the effectiveness of the proposed architectures, we plot detection error tradeoff (DET) curves for all comparable models, as illustrated in Figure \ref{fig:det}. We find that RSKNet-MTSP maintains a large
performance advantage across the entire operating range in all five test sets. RSKNet-MTSP-L is better than TDNN and E-TDNN, but worse than ResNet34-SP and ECAPA-TDNN for SITW dataset. For VoxCeleb dataset, RSKNet-MTSP-L performs better than ECAPA-TDNN in VoxCeleb1-E and VoxCeleb1-H across all operating range, and produces a similar curves with ECAPA-TDNN in VoxCeleb1-O.     

\begin{table}[htb]
    \centering
    \caption{EER for different durations on SITW.}
    \resizebox{\textwidth}{!}{
    \begin{tabular}{ccccccccc}
    \toprule
    \multirow{2}{*}{Model}  & \multicolumn{4}{c}{SITW-Dev} & \multicolumn{4}{c}{SITW-Eval}  \\
    \cline{2-9} 
    & $<$15s & 15-25s & 25-40s & $>$40s & $<$15s & 15-25s & 25-40s & $>$40s \\
    \midrule
    TDNN & 3.97 & 3.34 & 3.03 & 2.72  & 4.06 & 3.74  & 2.53  & 3.61  \\
    E-TDNN & 3.47 & 2.63 & 2.59  & 2.57  & 3.84  & 3.28  & 2.86  & 3.16 \\ 
    ECAPA-TDNN & 2.73 & \textbf{2.09} & 2.02  & 2.27  & 3.16 & 3.28  & 2.42 & 2.61 \\
    ResNet34-SP & 2.73 & 2.27 & 1.73  & 2.72  & 3.16 & 3.13  & 2.20  & 2.91  \\
    \textbf{RSKNet-MTSP} & \textbf{2.23} & 2.15 & \textbf{1.15}  & \textbf{2.12}  & \textbf{2.48} & \textbf{2.37}  & \textbf{1.98}  & \textbf{2.41}  \\
    \textbf{RSKNet-MTSP-L}  & 2.98  & 2.63 & 1.73 & 2.72 &  3.16 & 2.98 & 2.09 & 3.21 \\ 
    \bottomrule
    \end{tabular}
    }
    \label{tab:duration}
\end{table}

In order to verify the effectiveness of our proposed architectures on different durations, we divide the test utterances in SITW into four portions based on their durations and evaluate the performance on each part respectively. The EER results for different duration groups
on SITW is listed in Table \ref{tab:duration}. The first observation is that the group whose duration is less than 15s usually achieves the highest EERs than other groups, indicating that short utterances are more difficult for SV. The reason is that short utterances usually contain less speaker information than long utterances. 
In addition, RSKNet-MTSP achieves the lowest EERs in most duration groups except for 15s-25s in SITW-Dev, demonstrating the effectiveness of the proposed methods on both short and long utterances. We also find that RSKNet-MTSP produces similar EERs on $<$15s, 15-25s and $>$40s groups while the EERs of other models change a lot on different durations. This suggests that RSKNet-MTSP is robust with different durations.   

Next, we compare our models with recent state-of-the-art DNN-based TI-SV models. All models utilize the VoxCeleb1-O set as the test set and use different training sets. Table \ref{tab:vox1o} shows the EER results for various models. Note that prior studies \cite{8461375} indicate that the DNN architectures trained with a larger training set will obtain better performance. Both of the proposed RSKNet-MTSP and RSKNet-MTSP-L achieves competitive EER performance on the training set of the smallest VoxCeleb2 among all baseline models.



\begin{table}[htb]
    \centering
    \caption{Comparison with the state-of-the-art results on VoxCeleb1-O test set. The EERs of others models are the results listed in the references. $*$ denotes that training data are augmented with MUSAN \cite{snyder2015musan} and RIR \cite{7953152} datasets.} 
    \resizebox{\textwidth}{!}{
    \begin{tabular}{ccc}
    \toprule
    Model  &  Training set &  EER \\
    \midrule
    TDNN x-vector \cite{8461375} & VoxCeleb1+VoxCeleb2$^{*}$ & 3.10 \\
    TDNN+AMSoftmax+MHE \cite{Liu2019} & VoxCeleb1+VoxCeleb2$^{*}$ & 2.00 \\
    E-TDNN+AAMSoftmax \cite{zeinali2019description} & VoxCeleb2$^{*}$ & 1.46  \\
    DDB+Gate \cite{Jiang2019} & VoxCeleb1+VoxCeleb2$^{*}$ & 2.31 \\
    Thin-ResNet34+GhostVLAD \cite{NAGRANI2020101027} & VoxCeleb2 & 2.87 \\
    PRN-50v2+ft-CBAM+GhostVLAD \cite{9054440} & VoxCeleb2 & 2.03 \\
    ResNet34+LDE+FPMTC \cite{Jung2020} & VoxCeleb2 & 1.98 \\
    BLSTM-ResNet \cite{9053767} & VoxCeleb2 & 1.87 \\
    ResNet34+SP \cite{zeinali2019description} & VoxCeleb2$^{*}$ & 1.42  \\
    \textbf{RSKNet-MTSP}  &  VoxCeleb2 & \textbf{1.05}  \\
    \textbf{RSKNet-MTSP-L} & VoxCeleb2 & 1.68   \\
    \bottomrule
    \end{tabular}
    }
    \label{tab:vox1o}
\end{table}

\subsection{Ablation experiments}

In order to evaluate each component of the RSKNet-MTSP, we conduct several ablation experiments, where the results are shown in Table \ref{tab:ablation}. 

First, we evaluate the performance of SKConv modules on TI-SV tasks. The upper half of the Table \ref{tab:ablation} lists the EER results of different convolutions on VoxCeleb and SITW. DResNet34-1-MTSP adopts dilated convolutions with the rate of 2 in each convolutional layers of ResNet34.  DResNet34-2-MTSP performs dilated convolutions with the rate of 2 and 4 in convolutional layers of the 3rd and the 4th residual modules \cite{MIAO2021201}. The structure of DResNet34-1 and DResNet34-2 are displayed in Table \ref{tab:resnet}. We also construct the RSKNet-MTSP (w/o Att) architecture which does not employ the channel-wise attention part in SKConv modules. The results demonstrate that the dilated convolution does not yield a better performance, since dilated convolution ignores the neighbouring information of each CNN unit, leading to information loss in higher layers. The RSKNet-MTSP (w/o Att) achieves lower EER than ResNet-MTSP, which shows that the combination of standard convolution and dilated convolution can improve the performance of deep embedding architectures with the fusion of features by various spatial resolution. In addition, the RSKNet-MTSP (w/o Att) performs worse than RSKNet-MTSP, demonstrating that the channel-wise attention in SKConv can make the network focus on more important features for discriminating speakers. 

\begin{table}[htb]
    \centering
    \caption{The EER results of the ablation experiments.}
    \resizebox{\textwidth}{!}{
    \begin{tabular}{cccccc}
    \toprule
    Model  &  Vox1-O & Vox1-E & Vox1-H & SITW-D & SITW-E \\
    \midrule
    \textbf{RSKNet-MTSP} & \textbf{1.05} & \textbf{1.30} & \textbf{2.52} & \textbf{1.85} & \textbf{2.27} \\
    \midrule
    ResNet34-MTSP & 1.45 & 1.61 & 2.92 &  2.50  & 2.76 \\
    DResNet34-1-MTSP & 1.89 & 1.90 & 3.56 & 3.00  & 3.36 \\
    DResNet34-2-MTSP & 1.60 & 1.73 & 3.09 & 2.58   & 3.01 \\
    RSKNet-MTSP (w/o Att) & 1.26 & 1.38 & 2.55 & 2.04 & 2.46 \\ 
    \midrule
    RSKNet-SP & 1.41  &  1.50  & 2.71  & 2.12  & 2.57 \\
    RSKNet-GAP & 1.57 & 1.63 & 2.95 & 2.43  & 2.65 \\
    RSKNet-ASP & 1.37 & 1.52 & 2.72 & 2.04  & 2.32   \\
    RSKNet-MGAP & 1.54  & 1.54 & 2.83  & 2.12  & 2.49 \\
    \bottomrule
    \end{tabular}
    }
    \label{tab:ablation}
\end{table}

Secondly, we evaluate the effect of pooling methods on the proposed architectures. In addition to MTSP, we also implement statistics pooling (SP) \cite{Snyder2017}, global average pooling (GAP), attentive statistics pooling (ASP) \cite{Okabe2018} and multi-scale global average pooling (MGAP) \cite{Seo2019}. The results are shown in the bottom half of Table \ref{tab:ablation}. We find that RSKNet-GAP performs worse than RSKNet-SP in all test sets, which suggests that GAP loses some important speaker information. RSKNet-ASP achieves similar performance on VoxCeleb and better performance on SITW compared with RSKNet-SP. The reason is that most utterances in VoxCeleb are short utterances whose duration is less than 8s, where attention mechanism is hard to yield significant improvement. RSKNet-MGAP obtains lower EER than RSKNet-GAP, which also indicates that multi-scale aggregation captures more information for discriminating speakers based on their utterances.

\subsection{Evaluation of the lightweight architectures}

\begin{table}[htb]
    \centering
    \caption{The EER results for different kinds of convolutions.}
    \resizebox{\textwidth}{!}{
    \begin{tabular}{ccccccc}
    \toprule
    Model  & ParaM  & Vox1-O & Vox1-E & Vox1-H & SITW-D & SITW-E \\
    \midrule
    RSKNet-MTSP (SC)  & 13.9M & \textbf{1.05} & \textbf{1.30} & \textbf{2.52} & \textbf{1.85} & \textbf{2.27} \\
    RSKNet-MTSP (DSC) & 4.9M  & 1.45 & 1.60 & 3.00 & 2.27 & 2.84 \\
    RSKNet-MTSP (GC)  & 6.0M  & 1.41 & 1.58 & 2.95 & 2.35 & 2.82 \\
    ResNet34-SP (DSC) & 1.7M  & 1.78 & 1.85 & 3.12 & 2.77 & 3.39  \\
    ResNet34-SP (SC)  & 6.0M  & 1.43 & 1.59 & 2.76 & 2.27 & 2.82 \\
    \bottomrule
    \end{tabular}
    }
    \label{tab:dsc}
\end{table}

In this section, we evaluate the proposed lightweight architecture RSKNet-MTSP-L. We first compare various kinds of convolutions applied in the convolutional layers of RSKNet-MTSP, including standard convolution (SC), group convolutions (GC) and depthwise separable convolutions (DSC). The number of groups in GC is set to 4. Table \ref{tab:dsc} shows that RSKNet-MTSP (DSC) and RSKNet-MTSP (GC) perform worse than RSKNet-MTSP-SC, which indicates that DSC or GC reduce the performance of RSKNet-MTSP. Furthermore, RSKNet-MTSP (DSC) performs slightly worse than RSKNet-MTSP (GC) on VoxCeleb and SITW-Eval, but achieves better EER in SITW-Dev. Since the parameters of DSC are less than that of GC, we employ DSC in the convolutional layers of RSKNet-MTSP-L. Although DSC increases the EERs by 19\% to 38\% , it also reduces network parameters by 65\%, making the model portable. In addition, RSKNet-MTSP (DSC) yields similar EERs with ResNet34-SP (SC) in four test sets except for VoxCeleb1-H, with only 81.7\% network parameters of ResNet34-SP (SC). On the other hand, when utilizing the DSC, the RSKNet-MTSP (DSC) significantly outperforms ResNet34-SP (DSC), demonstrating the effectiveness of RSKNet-MTSP.

\begin{table}[htb]
    \centering
    \caption{The EER results for different values of the low-rank constant $p$. ``w/o" denotes the normal model without low-rank factorization.}
    \resizebox{\textwidth}{!}{
    \begin{tabular}{cccccccc}
    \toprule
    Model & $p$ & ParaM & Vox1-O & Vox1-E & Vox1-H & SITW-D & SITW-E \\
    \midrule
    \multirow{4}{*}{RSKNet-MTSP-L} 
    & w/o & 4.9M & \textbf{1.45} & \textbf{1.60} & \textbf{3.00} & \textbf{2.27} & \textbf{2.84} \\
    & 100 & 3.3M & 1.70 & 1.75 & 3.15 & 2.54 & 3.25 \\
    & 150 & 3.8M & 1.68 & 1.71 & 3.01 & 2.46 & \textbf{2.84} \\ 
    & 200 & 4.3M & 1.76 & 1.78 & 3.08 & 2.89 & 3.06 \\
    \midrule
    \multirow{4}{*}{RSKNet-MTSP}
    & w/o & 13.9M & \textbf{1.05} & \textbf{1.30} & 2.52 & 1.85 & 2.27 \\ 
    & 100 & 12.3M & 1.14 & 1.32 & \textbf{2.44} & \textbf{1.73} & \textbf{2.08} \\
    & 150 & 12.9M & 1.33 & 1.38 & 2.56 & 1.93 & 2.35 \\
    & 200 & 13.4M & 1.27 & 1.36 & 2.48 & 1.89 & 2.35 \\
    \bottomrule
    \end{tabular}
    }
    \label{tab:lowrank}
\end{table}

We then discuss the results of models with low-rank factorization. For a more comprehensive comparison, we apply low-rank technique as stated in Section \ref{sec:lowrank} to both RSKNet-MTSP-L and RSKNet-MTSP. Note that all models are initialized by random initialization and then trained without any fine-tuned training. The low-rank constant $p$ is set to 100, 150 and 200, respectively. Table \ref{tab:lowrank} shows the EER results of different models. The low-rank factorization technique makes a different impact on RSKNet-MTSP and RSKNet-MTSP-L. For RSKNet-MTSP-L, we find that the low-rank models significantly reduce the network parameters and they perform slightly worse than the normal model. When p is set to 150, RSKNet-MTSP-L achieves the best performance for all test sets in low-rank models. 
Compared with the normal model, RSKNet-MTSP-L ($p=150$) yields similar EERs for Vox1-H and SITW-E with 22.4\% less parameters. For RSKNet-MTSP, it can be observed that the low-rank model outperforms the normal model in three tasks when $p$ is set to 100, which indicates that the low-rank matrix factorization may contribute to better performance for SV tasks. 

\begin{table}[htb]
    \centering
    \caption{The EER results for different models.}
    \resizebox{\textwidth}{!}{
    \begin{tabular}{ccccccc}
    \toprule
    Model & ParaM & Vox1-O & Vox1-E & Vox1-H & SITW-D & SITW-E \\
    \midrule
    RSKNet-MTSP-L & 3.8M & \textbf{1.68} & \textbf{1.71} & \textbf{3.01} & \textbf{2.46} & \textbf{2.84} \\
    lite RSKNet-MTSP & 3.5M & 1.70 & 1.76 & 3.27 & 2.58 & 3.03 \\
    \bottomrule
    \end{tabular}
    }
    \label{tab:lite}
\end{table}

Another method to reduce network parameters is to directly reduce the channel dimension of the network. To compare these two methods, we also design a lite RSKNet-MTSP model, which halves the channel dimension of each convolutional layer in RSKNet-MTSP. The EER results shown in Table \ref{tab:lite} show that RSKNet-MTSP-L achieves lower EERs than lite RSKNet-MTSP in all test sets with similar model parameters, further demonstrating the effectiveness of RSKNet-MTSP-L.

\section{Conclusion}

\label{sec:conclusion}

In this paper, we propose a novel CNN-based deep embedding architecture RSKNet-MTSP for SV, where advanced techniques such as residual selective kernel blocks (RSKBlocks) and  multiple time-scale statistics pooling (MTSP) module are introduced. The RSKBlock captures both long temporal context and neighbouring information, and applies a channel-wise attention mechanism to adaptively adjust the weight between short and long context of input utterances. The MTSP module extracts speaker features in terms of temporal variations over long-term context from multi-scale features of the network, reducing the information loss of the CNN model. We then propose a lightweight architecture based on RSKNet-MTSP, called RSKNet-MTSP-L, for real applications with limited resources. RSKNet-MTSP-L employs depthwise separable convolution and low-rank factorization to reduce the network parameters for a portable model.

The two proposed models are evaluated on two well-known public datasets, VoxCeleb and Speaker in the Wild (SITW). From the experimental results, the RSKNet-MTSP achieves at least 9\%-26\% improvement over the state-of-the-art models in all five test sets; while the RSKNet-MTSP-L achieves competitive performance with less network parameters. The ablation experiments further indicate that the proposed approaches substantially improve the deep embedding architectures for SV tasks. In the future, we aims to  investigate more advanced compression methods and make the model more portable.

\section*{Acknowledgement}

This work is supported by National Natural Science Foundation of China (Grant No. 62002177), Science and Technology Planning Project of Tianjin, China (Grant No. 20YDTPJC01810), Tianjin Natural Science Foundation (Grant No. 19JCQNJC00300) and Research and Innovation Project for Postgraduates in Tianjin (Artificial Intelligence) (Grant No. 2020YJSZXB01).

\bibliography{mybibfile}

\end{document}